\documentclass[runningheads]{llncs}
\usepackage[T1]{fontenc}
\usepackage{stmaryrd}
\usepackage[textwidth=3cm,tickmarkheight=3pt,disable]{todonotes}
\usepackage{xspace}
\usepackage{amsmath}
\usepackage{amssymb}
\usepackage{mathpartir}
\usepackage[dvipsnames]{xcolor}
\usepackage[colorlinks=true, linktoc=all,
    linkcolor=Blue,
    citecolor=Blue,
    filecolor=magenta,
    urlcolor=Fuchsia,
    pdftitle={Case Study: Saturations as Explicit Models in Equational Theories},
    pdfauthor={Janota, Rawson, Schulz}]{hyperref}

\usepackage{orcidlink}
\renewcommand{\orcidID}[1]{\orcidlink{#1}}

\newcommand{\eq}{\approx}
\newcommand\nteq{\not\approx}
\newcommand{\normalise}[1]{\llbracket #1 \rrbracket}
\newcommand{\vampire}{\textsc{Vampire}\xspace}
\usepackage{numprint}
\usepackage{booktabs}
\npthousandsep{,}

\newcommand\isafor{%
  \textsf{Isa\kern-0.15exF\kern-0.15exo\kern-0.15exR}%
  \xspace%
}
\newcommand\ceta{%
  \textsf{C\kern-0.15exe\kern-0.45exT\kern-0.45exA}%
  \xspace%
}
\newcommand\CSI{%
 \textsf{C\kern-0.05emS\kern-0.05emI}%
 \xspace
}
\newcommand\TTTT{%
 \textsf{T\kern-0.15em\raisebox{-0.55ex}T\kern-0.15emT\kern-0.15em\raisebox{-0.55ex}2}%
 \xspace%
}

\begin{document}
\title{Case Study: Saturations as Explicit Models in Equational Theories}
\author{
Mikol\'a\v{s} Janota\inst{1}\orcidID{0000-0003-3487-784X} \and
Michael Rawson\inst{2}\orcidID{0000-0001-7834-1567} \and
Stephan Schulz\inst{3}\orcidID{0000-0001-6262-8555}
}
\authorrunning{Janota et al.}

\institute{
Czech Technical University in Prague, CIIRC, Czechia \email{mikolas.janota@cvut.cz}\and
University of Southampton, UK \email{michael@rawsons.uk}\and
DHBW Stuttgart, Stuttgart, Germany \email{schulz@eprover.org}
}

\maketitle
\begin{abstract}
  Automated theorem provers (ATPs) can disprove conjectures by
  saturating a set of clauses, but the resulting saturated sets are
  opaque certificates.  In the unit equational fragment, a saturated
  set can in fact be read as a convergent rewrite system defining an
  explicit, possibly infinite, model --- but this is not widely known,
  even amongst frequent users of ATPs.  Moreover, ATPs do not emit
  these explicit certificates for infinite (counter-)models. We
  present such a certificate construction in full, implement it in
  \vampire and E, and apply it to the recent Equational Theories
  Project~\cite{etp}, where hundreds of implications do not admit
  finite countermodels.  The resulting rewrite systems can be checked
  for confluence and termination by existing certified tools, yielding
  trustworthy countermodels.

\end{abstract}

\section{Introduction}
In a September 2024 post on his blog~\cite{blog}, Terence Tao proposed
\emph{The Equational Theories Project} (ETP), a collaborative project
bringing together mathematicians and automated reasoning
researchers. The goal was to classify a set of~\numprint{22028942}
implications between universally-quantified equalities over a single
binary operation (see Section~\ref{sec:study} for details). Each such
implication yields a formula that can be processed by a first-order
automated theorem prover (ATP). An experimental study shows that ATPs
are very successful on such problems~\cite{janota25experiments}.  From
the initial 22 million, fewer than 1000 are left unsolved by
\vampire~\cite{vampire}.

The ETP is now completed~\cite{etp}, but at considerable human
cost: the report counts 34 authors. A significant obstacle in using
modern ATPs is that the mathematicians are also interested in
\emph{why} an implication holds or does not: a certificate.  If the
implication in question holds, ATPs are capable of producing a
detailed proof. However, if the implication does not hold, a
\emph{counter-model} is required. Current ATPs cannot produce an
explicit counter-model, but they can sometimes provide a
\emph{saturation} that implicitly describes a model. \emph{Finite}
counter-models can be obtained by the use of a \emph{finite model
  builder}~\cite{paradox,fmb}.  However, in some cases all models are infinite. Moreover,
large finite models are often out of reach --- in fact,
large finite models can be harder to find than the infinite ones
represented by saturations. This is due to the combinatorics of known
finite model building algorithms, which typically become too large for
domain sizes above around 20.  However, nontrivial infinite (counter-)
models are often needed for classification of algebraic
structures~\cite{Phillips2005,Kinyon2013}.  This motivates our case
study, where we modify current tools to produce
such models.

Saturation-based ATPs such as E~\cite{SCV:CADE-2019},
\vampire~\cite{vampire}, iProver~\cite{DK:IJCAR-2020},
Zipperposition~\cite{VBBCNT:JAR-2022},
Twee~\cite{Smallbone:CADE-2021}, or Waldmeister~\cite{Hillenbrand2002}
search for a refutation by deducing consequences of the input
formulae. If falsum is derived, the input was unsatisfiable. Another
possibility is that \emph{saturation} is achieved: no more new
non-redundant consequences can be derived. If the proof procedure used
is \emph{complete} --- that is, a proof of falsum is guaranteed to be
found whenever the input is unsatisfiable --- by contraposition the
input is satisfiable. Typical ATP systems at best dump the saturated
set of consequences to report satisfiability.

Saturated sets are unwieldy objects. Unlike proofs, they are
opaque and rarely useful as counterexamples in downstream applications such as mathematics or
verification. This problem has become more acute as ATP systems
improve and saturate more often. Recent calculus improvements such as
ground joinability~\cite{ground-joinability} and AVATAR~\cite{AVATAR},
combined with aggressive preprocessing such as blocked clause
elimination~\cite{bce} mean that systems saturate on more inputs than
ever before, including some traditionally-difficult inputs like
associative and commutative operators.

We would prefer to extract a more explicit representation of a model
from saturations, one which allows at least evaluating ground
terms. Doing this for full clausal logic is very
difficult~\cite{saturation-based-model-building}, but for certain
fragments it is considerably easier. We consider the \emph{unit
  equational} fragment, which allows only universally-quantified
equations and ground disequations. This fragment is nonetheless very
powerful: many algebraic structures can be expressed as sets of unit
equations\footnote{``admits a unit equational axiomatisation'' is
  nearly the definition of a \emph{variety}}; it suffices for many
applications in functional programming; and even full first-order
logic can be embedded in it, the Horn fragment very
efficiently~\cite{horn2ueq}.

For the unit equational fragment, there is a straightforward reading
of saturated sets as potentially-infinite models
(Section~\ref{sec:construction}), yet this reading is underexploited
in practice --- in part because it relies on subtleties
of rewriting theory that are not obvious to end-users.
We aim to bridge this gap, making saturations
directly usable as counter-models for non-specialists,
in particular mathematicians.

\section{Preliminaries}\label{sec:prelim}
We assume familiarity with classical first-order logic~\cite{smullyan}
with equality.  ATP systems take a set of \emph{axiom} formulas
$\Gamma$ and a single conjecture formula $C$, and try to prove or
disprove $C$ from $\Gamma$, completely automatically and without human
intervention.  In the unit equational case, $C$ and every formula in
$\Gamma$ are of the form $\forall x_1\ldots x_n.~s \eq t$.  ATPs
typically proceed by refutation, negating $C$ and trying to reach a
contradiction.  As $C$ is negated, the negation is pushed inwards and
$C$ becomes a disequation $s \not\eq t$, containing no variables after
Skolemisation.

The initial set of clauses $E$ is therefore $\Gamma, s \not\eq t$.
Saturation-based ATPs then exhaustively apply inferences of a complete
calculus to $E$, adding any consequences back to $E$.  If a
contradiction is produced, $C$ is a theorem of $\Gamma$.  However, if
no contradiction is detected and no non-redundant new clauses can be
derived that are not already in $E$, then $E$ is \emph{saturated} and
$\Gamma, s \not\eq t$ is \emph{satisfiable}.  Otherwise, a proof would
have been found by the complete calculus.  Therefore there is a model
(or counterexample) satisfying $\Gamma, s \not\eq t$, which shows
that $C$ is not a theorem of $\Gamma$.  In
Section~\ref{sec:construction} we show how to construct this
implicitly-represented model for saturation-based ATPs that use
calculi descended from \emph{completion}~\cite{KB70}.

The \emph{signature} of a unit equational problem is the set of
functions and constants in $\Gamma, s \not\eq t$.  A term or equation
is \emph{ground} if it contains no variables.  We write $s[t]$ to
indicate that $t$ is a sub-term of $s$, not excluding the possibility
that $s$ \emph{is} $t$.  The \emph{Herbrand universe} is the set of
all ground terms that can be constructed from symbols in the
signature.  Substitutions $\sigma$ are mappings from variables to
terms.  Their application to a term, $\sigma(t)$, replaces every variable
$x$ in $t$ with $\sigma(x)$.
Equations are considered implicitly symmetric.

\section{Constructing Explicit Models}
\label{sec:construction}

Suppose we are given a saturated set $E$ of universally-quantified
equations\footnote{Disequations in the saturation can be ignored for
  our purposes.} from an ATP based on unfailing
completion~\cite{HR87,BDP89} or superposition~\cite{BG94}. We are also
given the signature, and details of its internal term ordering
$\prec$, which is total, well-founded, and decidable on ground terms.
We will now explicitly construct an equality Herbrand model for $E$
under $\prec$ such that ground terms and equations can be evaluated.

The domain $D$ is the Herbrand universe of the signature, i.e.\ ground
terms interpreted only as themselves. We therefore need only define
the equality relation $\eq$. The idea is that terms will be equal
when they have the same normal form when rewritten by $E$. To this
end, we define a computable normalisation function
$\normalise{\_} : D \to D$.

\begin{definition}[rewrite]
  A ground term $t[\sigma(l)]$ can be rewritten to $t[\sigma(r)]$ when
  there is a ground instance $\sigma(l) \eq \sigma(r)$ of an equation
  $l \eq r \in E$ with $\sigma(l) \succ \sigma(r)$.
\end{definition}
Note that $l \succ r$ does not necessarily hold when rewriting, although it does in
the case that $\sigma(l) = t$, because $\succ$ is total on ground
terms. Finding a rewrite can be done algorithmically by matching
either side of an equation onto any subterm of $t$, then checking the
ordering constraint. If there are unbound variables on the other side
of the equation, they can be mapped to any ground term such that $\sigma(l) \succ \sigma(r)$, in
practice the smallest constant.

\begin{definition}[normalisation]
  The normal form $\normalise{t}$ of a term $t$ under $E$ is the result of
  rewriting $t$ until fixed point.
\end{definition}
Any such computation terminates, since $\prec$ is
well-founded. Additionally, the precise order of rewrites found by a
particular normalisation function does not matter, since the set of
equations resulting if the unfailing completion procedure terminates
is confluent on ground terms~\cite{BDP89}.

\begin{example}
  Consider a binary function $f$, and the lexicographic path ordering
  with $f \succ b \succ a$. $f$ is commutative, and this results in
  the saturated set $f(x, y) \eq f(y, x)$. One possible computation
  normalising $f(f(b, a), a)$ is
$$
\normalise{f(f(b, a), a)} =
\normalise{f(f(a, b), a)} =
\normalise{f(a, f(a, b))} =
f(a, f(a, b))
$$
\end{example}
We can now define $s \eq t$ to be the equivalence relation
$\{(s, t)~|~\normalise{s} = \normalise{t}\}$ and we have our model. It
remains to show that this is a model of the input. The input
disequation $s \not\eq t$ has $\normalise{s} \neq \normalise{t}$,
otherwise the ATP would have derived falsum. It is therefore
satisfied. Input equations $l \eq r$ are true by assumption in the
original equational theory, which is invariant under (unfailing)
completion. Hence there is a proof of $l \eq r$ in the saturated
system, and by the equivalence of Church-Rosser and confluence, a
rewrite proof for any ground instance of it. So,
$\normalise{\sigma(l)} = \normalise{\sigma(r)}$ for any grounding
substitution $\sigma$.

Ground terms can be easily evaluated as their normal forms in such
models, and ground equations are also decided by normalisation. We
remark that each saturated clause set and $\prec$ induces a unique model. A
limitation of this approach is that there is no obvious way to
determine whether two models are equivalent, nor to evaluate
non-ground terms or non-ground (dis)equations.

\section{Case Study: Countermodels of Equational Theories}
\label{sec:study}

The ETP targets
implications between universally-quantified equalities over a single
binary operation~$*$, which we motivate by the following examples.
\begin{align}
	(x*y)*z\eq x*(y*z) &\implies x*y\eq y*x\label{eq:1}\\
	x*y\eq y*x &\implies (x*y)*z\eq x*(y*z)\label{eq:2}\\
	x*y\eq u*w &\implies x*y\eq y*x\label{eq:3}
\end{align}
Implication~\eqref{eq:1} asks whether associativity implies
commutativity; it does not hold, e.g., matrix multiplication is
associative but not commutative. Likewise,~\eqref{eq:2} asks whether
commutativity implies associativity; this fails because, e.g.,
$\frac{x+y}{2}$ is commutative but not associative. In contrast,
implication~\eqref{eq:3} holds: the left-hand side forces~$*$ to
be a constant function, which is necessarily commutative. The ETP
considered all equalities where the operation $*$ is applied at most
4~times. There are $n=\numprint{4694}$ equations, systematically
numbered from \texttt{1} to \texttt{4694}, yielding
$n^2-n = \numprint{22028942}$ possible implications. The ETP has
successfully decided all the considered implications, and the results
can be explored on the project's website~\cite{dashboard}.

The previous experiment~\cite{janota25experiments} by Janota shows that
\vampire can prove \emph{all} implications that hold. These already
have a certificate in the form of a superposition proof.
On the other hand, \numprint{13854015} problems were shown \emph{not} to hold by \vampire's
finite model builder (FMB)~\cite{paradox,fmb}, which produces a checkable certificate.
A further \numprint{817} non-theorems were detected by saturation.
We ran the finite model builder on these \numprint{817} with a longer timeout, leaving us with 304 non-theorems refuted only by saturation.
Of these, 196 are marked \emph{finitely unsatisfiable} by Infinox~\cite{infinox}, so no finite counter-model exists.
This means that the remaining 108 \emph{may} have a finite counter-model, but we were not able to obtain it via FMB.
A detailed table of the experiments is available on the authors'
website\footnote{\url{https://people.ciirc.cvut.cz/~janotmik/stamp}}.
Further reductions are possible by identifying \emph{duals}: implications that
can be obtained from one another by swapping order of arguments, which we avoid here for clarity.
The numbers are summarized in the following table.
\begin{center}
  \setlength{\tabcolsep}{10pt}
  \begin{tabular}{r ccc}
    \toprule
              & $\neg$Fin.\ Unsat & Fin.\ Unsat & Total \\
    \midrule
    $\neg$FMB & 108               & 196         & \textbf{304}   \\
    FMB       & 513               & 0           & 513   \\
  \end{tabular}
\end{center}

\begin{figure}[t]
  \begin{center}
\begin{equation*}
\begin{aligned}[t]
f_2 * f_4 &\rightarrow f_0(b,f_4) \\
f_0(f_0(f_0(x,x),x),x) &\rightarrow x \\
f_2 * a &\rightarrow f_0(b,a) \\
f_0(f_2,b) &\rightarrow f_4 \\
y * f_0(x,y) &\rightarrow x \\
f_0(x,f_0(x,x)) &\rightarrow x \\
x * f_0(x,y) &\rightarrow f_0(y,f_0(x,y)) \\
(x * y) * y &\rightarrow f_0(x,y) \\
\end{aligned}
\qquad
\begin{aligned}[t]
f_4 * b &\rightarrow f_0(f_3,b) \\
b * a &\rightarrow f_2 \\
f_3 * b &\rightarrow f_4 \\
f_1(y,x) &\rightarrow x \\
f_2 * b &\rightarrow f_3 \\
f_0(x,y) * y &\rightarrow f_0(x * y, y) \\
b * f_4 &\rightarrow f_2 \\
x * x &\rightarrow f_0(f_0(x,x),x) \\
\end{aligned}
\end{equation*}
   \end{center}
  \caption{Confluent and terminating system for $118\land \lnot 274$}\label{fig:rw}
\end{figure}
As an example, consider disproving the implication
between equations \texttt{118} and \texttt{274}: $ \forall xy.\,x \eq y((xy)y)\implies \forall xy.\, x \eq ((yx)y)y $.
After negation and Skolemization, we obtain the input clause set
\begin{mathpar}
  x \eq y((xy)y)\and
  a \nteq ((ba)b)b
\end{mathpar}
which is satisfiable but admits only infinite models.\footnote{For the interested reader, this is shown in Appendix~\ref{funsat}.}
The rewrite system for the counter-model is shown in Fig.~\ref{fig:rw}.
Note that \vampire has introduced definitions
\begin{mathpar}
f_0(x, y) \triangleq (x * y) * y\and
f_1(x, y) \triangleq x * f_0(y, x)\\
f_2 \triangleq b * a\and
f_3 \triangleq f_2 * b\and
f_4 \triangleq f_3 * b
\end{mathpar}
in order to achieve saturation.
These definitions cannot be easily read off the saturated set, but they are not necessary to define the model.
Perhaps surprisingly, in some cases the original problem is already saturated.
This is the case for implication between \texttt{477} and \texttt{1426},
which also has only infinite counter-models.\footnote{by an identical argument}
\begin{mathpar}
x \eq (y (x (y (y y))))\and
a \nteq (a a) (a (a a))
\end{mathpar}
Here the positive equation is pre-oriented right-to-left, and no consequences are produced from either clause in the calculus.

\section{Checked Models}
Ideally, models would be externally checkable.  Systems like E or
\vampire are large, complex pieces of software, and even their
underlying theory can be tricky~\cite{bachmair-revisited}.  One path
to more trust we envisage is exporting a model suitable for checking
in a proof assistant, such as the Lean formalisation employed in the
ETP.  This would require:
\begin{enumerate}
\item Defining the domain as a Herbrand universe. This is a
  straightforward inductive definition in most systems.
\item Defining a normalisation function. Many systems require that all
  user-defined functions terminate, which typically requires showing
  that $\prec$ is well-founded. This may not be trivial, and depends
  on fine details of the term ordering.
\item Showing that the resulting model is a model of the input. This
  should be trivial for ground disequations, but
  universally-quantified equations will require a more involved
  argument.
\item Therefore, we will also very likely need to show confluence of
  the rewrite system. This is guaranteed in theory but will need to be
  re-proven, perhaps using a confluence checker.
\end{enumerate}
Such a system would be quite involved.  However, to demonstrate the
concept, we have modified \vampire and E to print the rewrite system on
saturation, provided that every equation $l \eq r$ is ``pre-ordered'',
i.e.\ $l \succ r$.
Of the 304 saturations of interest, to our surprise 261 are
pre-ordered.  We used the tool \CSI~\cite{Nagele2017,Zankl2011} to
check for (ground) confluence of the rewrite system, and
\TTTT~\cite{ttt2} for termination. Both tools produce certificates
independently verifiable by the tool \ceta~\cite{Thiemann2009}, which
itself relies on the \isafor Isabelle library.  All 261 pre-ordered rewrite
systems were certified as confluent and terminating by \CSI and \TTTT.
In the other 43 cases, a model can still be constructed, but it is not
yet clear to us how to automatically check (ground-) confluence and
termination.

\section{Related Work}\label{sec:rw}
Bachmair-Ganzinger model construction~\cite{BG94}
builds a Herbrand model from a saturated set by
generating rewrite rules from so-called \emph{productive} clauses.
More recently, Lynch~\cite{lynch13}
revisits model construction, giving a more explicit and algorithmic
account. The construction in Section~\ref{sec:construction} is a
specialisation of this framework to the unit equational fragment,
where the saturated set of equations directly forms a convergent
ground rewrite system and the model is the quotient of the Herbrand
universe by normalisation.
Unit equations allow us to dispense with clause-level
machinery and use results from unfailing completion~\cite{BDP89} directly.

Peltier~\cite{peltier-jsc03} develops methods for extracting Herbrand
models from clause sets saturated under ordered resolution,
and extends them to refinements such as hyperresolution~\cite{peltier-ic03},
as do Ferm\"uller and Leitsch~\cite{hyperresolution}.
Peltier also builds infinite models from equational saturated sets,
defining \emph{non-ambiguous} formulae such that
model construction is well-defined~\cite{peltier03}.
In our unit equational setting,
non-ambiguity is guaranteed by the convergence of the rewrite system
obtained from saturation.
Horbach and Weidenbach study
decidability results for saturation-based model building in full
first-order logic~\cite{saturation-based-model-building}.
Infinite models can also be constructed by looking for a model of a fixed
template~\cite{BrownJK22,elad2024infinite}, by identifying specific
forms of the formula~\cite{ge-cav09}, or by enumeration~\cite{ParsertBJK23}.

The TSTP format for ATP solutions has recently been extended~\cite{tstp,SteenSFM23}
to allow for reporting both finite and infinite models.
In principle, the models we construct could be reported in this format.

\section{Summary and Future Work}\label{sec:future}
We have shown that saturated sets produced by ATPs in the unit
equational fragment can be read as (ground-)convergent rewrite systems
defining explicit, potentially infinite models.  The construction
relies on known results in rewriting but has not been widely
accessible to users of ATPs; we present it here in detail. Applying
the construction to the Equational Theories Project, we modified
the saturation-based ATPs \vampire and E to
emit models in the form of ground-convergent rewrite systems, even for
problems where no finite counter-model exists. We demonstrated that
all resulting saturations with pre-ordered equations can be verified
for confluence and termination by certified tools, yielding 261 fully
verified counter-models. This independent verification also corroborates that the
ATPs emit correct saturations, even for the 43 problems with
unoriented equations, which are currently beyond our verification approach.

Extending the approach beyond the unit equational fragment to richer
clausal logics is a challenging but important goal. At least for the
Horn case, this should be possible either by embedding Horn problems
into UEQ~\cite{horn2ueq}, or by treating the axioms as conditional
rewrite rules in the style of Dershowitz~\cite{Dershowitz:IJCAI-91}. A further question is whether (potentially large)
\emph{finite} models could be constructed explicitly from a convergent
rewrite system.

\subsubsection{\discintname}
The authors have no competing interests to declare that are
relevant to the content of this article.
%

\bibliographystyle{splncs04}

\appendix
\section{No finite counterexample to $\mathtt{118} \implies \mathtt{274}$}
\label{funsat}
The following clause set has no finite model.
\begin{equation}\label{eq:axiom}
x \eq y * (( x * y) * y)
\end{equation}
\begin{equation}\label{eq:goal}
a \nteq (((b * a) * b) * b)
\end{equation}

\begin{proof}
Suppose that there is a finite model for contradiction.
First, observe that left-multiplication is surjective: by \eqref{eq:axiom} every $x$ is of the form $y * z$, where $z$ is $(x * y) * y$.
Surjective functions on finite sets are injective.
Now consider the following instance of \eqref{eq:axiom}:
$$
b * a \eq b * (((b * a) * b) * b),
$$
and obtain
$$
a \eq ((b * a) * b) * b,
$$
by injectivity.
\qed
\end{proof}

\end{document}